\documentclass[twocolumn,showpacs,preprintnumbers,amssymb,prl]{revtex4-1}
\usepackage{pgfplots}
\usepackage{latexsym,epsfig}
\usepackage{graphicx}
\usepackage{dcolumn}
\usepackage[normalem]{ulem}
\usepackage{amsthm,amsmath}

\begin{document}
\title{Relativistic Normal Coupled-Cluster Theory for Accurate Determination of  Electric Dipole Moments of Atoms: First application to $^{199}$Hg atom}

\author{B. K. Sahoo}
\email{bijaya@prl.res.in}
\affiliation{Atomic, Molecular and Optical Physics Division, Physical Research Laboratory, Navrangpura, Ahmedabad-380009, India}
\affiliation{State Key Laboratory of Magnetic Resonance and Atomic and Molecular Physics, Wuhan Institute of Physics and Mathematics
Chinese Academy of Sciences, Wuhan 430071, China}

\author{B. P. Das}
\email{bpdas.iia@gmail.com}
\affiliation{Department of Physics and International Education and Research Center of Science, Tokyo Institute of Technology, 2-12-1 Ookayama Meguro-ku, Tokyo 152-8550, Japan}

\date{Received date; Accepted date}

\begin{abstract}
Recent relativistic coupled-cluster (RCC) calculations of electric dipole moments (EDMs) of diamagnetic atoms due to parity and time-reversal violating (P,T-odd) interactions,
that are essential ingredients for probing new physics beyond the standard model of particle interactions, differ substantially from the previous theoretical results. It is 
therefore necessary to perform an independent test of the validity of these results. In view of this, the normal coupled-cluster (NCC) method has been extended to the 
relativistic regime to calculate the EDMs of atoms by simultaneously incorporating the electrostatic and P,T-odd interactions in order to overcome the shortcomings of the
ordinary RCC method. This new relativistic method has been applied to $^{199}$Hg, which currently has a lower EDM limit than that of any other system. The results of our 
relativistic NCC and self-consistent RCC calculations of the EDM of this atom are found to be close. The discrepancies between these two results
on the one hand and those of previous calculations on the other are elucidated. Furthermore, the electric dipole polarizability of this atom, which has computational 
similarities with the EDM, is evaluated and it is in very good agreement with its measured value.
\end{abstract}

\maketitle
 
The combined results of experiments and relativistic many-body calculations for atomic electric dipole moments (EDMs) due to parity and time-reversal violating (P,T-odd) 
interactions can provide important insights into CP violation corresponding to mass scales of tens of TeV or larger \cite{barr,pospelovreview} and thereby
probe new physics beyond the standard model (BSM). Of all the systems on which experiments have been performed, mercury yields the lowest EDM limit to date. The measured value
of the EDM for $^{199}$Hg, $d( ^{199}\rm{Hg})=(-2.20 \pm 2.75(\rm{stat}) \pm 1.48 (\rm{sys})) \times 10^{-30}$ $e$cm, translates to an upper limit of $|d( ^{199}\rm{Hg})| 
< 7.4 \times 10^{-30}$ $e$cm with 95\% confidence level \cite{graner}. Further improvement of this limit by 2-3 orders  could provide crucial information on the validity of certain 
BSMs, in particular some variants of the multi-Higgs, supersymmetric, left-right symmetric models \cite{barr,pospelovreview,yamanaka}. It would, therefore, be very desirable to
determine precise limits for CP violating parameters from the EDM of $^{199}$Hg. Rigorous calculations of EDMs considering the nuclear Schiff moment (NSM) and the electron-nucleus 
tensor-pseudotensor (T-PT) interactions in this atom have been performed using the relativistic coupled-cluster (RCC) theory in the singles and doubles approximation (CCSD method) 
\cite{yamanaka,yashpal,bksahoo}. By combining the results of the latest two calculations \cite{yamanaka,bksahoo} with that of the measured EDM value of $^{199}$Hg,
we had obtained upper limits for the NSM, $|S| < 4.2 \times 10^{-13} |e| \rm{fm}^3$, and the T-PT coupling coefficient, $|C_T| < \ 7.0 \times 10^{-10}$. Further combining these 
values with the nuclear calculations and quantum chromodynamics (QCD), gave the limits $d_n  <  2.2 \times 10^{-26} |e| \rm{cm}$ and $d_p <  2.1 \times 10^{-25} |e| \rm{cm}$ 
for the EDMs of the neutron and the proton respectively, and  $|\bar{\theta}| < 1.1 \times 10^{-10}$ and $|\tilde{d}_u - \tilde{d}_d| < 5.5 \times 10^{-27} |e|\rm{cm} $ for the QCD 
parameter and the combined up- and down- quark chromo-EDMs respectively \cite{yamanaka,bksahoo}. 

 Our recent RCC results  \cite{yamanaka,bksahoo} differ by about 20-50\% from those obtained previously by employing a variety of relativistic many-body methods (discussions on
these results can be found in Refs. \cite{yamanaka,yashpal}). The results of another RCC calculation, referred to in the literature as the perturbed RCC (PRCC) method,
for $^{199}$Hg co-authored by one of us \cite{latha} is also not in agreement with our latest work \cite{bksahoo} on this subject. The differences between the two RCC methods are
essentially technical in nature. The PRCC approach accounts for the correction from the normalization of the wave function and allows disconnected terms appearing in the numerator
of the expressions in the calculation of the EDM of a closed-shell atom \cite{latha,mani,angom}. In our formalism of the RCC method, the expectation value of an operator contains 
only connected terms in the numerator after the normalization cancels out in the numerator and the denominator \cite{cizek,bartlett}. In one of the RCC calculations of the
properties of $^{199}$Hg, a large number of lower-order terms arising from the non-terminating connected terms of the exponential terms containing the cluster excitation and de-excitation 
operators in the expression for the EDM  have been computed with relatively less computational effort \cite{yashpal}. The results of these properties were in reasonable agreement with some
of the results reported previously, but further attempts to include higher-order non-linear terms through an iterative procedure reduced the magnitudes of the results significantly
\cite{yamanaka,bksahoo}. In fact, the discrepancies between these results and those of previous calculations are about 50\% in the case of  EDMs for some of the  other atoms that 
are of experimental interest such as $^{171}$Yb \cite{sahoo} and $^{225}$Ra \cite{singh}. Thus, it is imperative to test the validity of these calculations by taking recourse to a reliable 
method that is capable of overcoming the drawbacks of the regular RCC method. 
 
 The coupled-cluster (CC) method has been applied to atoms \cite{bartlett,deyonker,nataraj,pathak,eliav}, molecules \cite{cizek,bartlett,crawford,prasana}, condensed
matter systems \cite{bishop01,bishop02,bishop} and nuclei \cite{kowalski,hagen}. It is currently one of the leading quantum many-body methods and has been referred 
to as the gold standard for treating electron correlation \cite{bartlett,crawford,lindgren,bishopbook1,bishopbook}. It is straightforward to apply  RCC methods for the evaluation of energies,
but not for other properties. The normal coupled-cluster (NCC) method, in which the bra and ket satisfy the bi-orthogonal condition, is tailor-made for the evaluation of expectation values of operators 
corresponding to different properties \cite{bishop,bishopbook1,bishopbook,bishop1,arponen}. This method, unlike the usual RCC method possesses two important attributes which
makes it attractive for the calculations of different atomic properties. The first being that it satisfies the Hellman-Feyman theorem and the second is that expectation values of 
operators terminate in a natural way \cite{bishop,bishop1,arponen}. Thus the development of the NCC method in the relativistic framework (RNCC) can lead to an improvement in 
the accuracies of the calculations of atomic properties. Unlike molecular calculations, it is possible to exploit the spherical symmetry of the systems and to treat parity as a good
quantum number to develop the RNCC method for atoms in the presence of P,T-odd interactions. As these interactions are very weak compared to the electrostatic interactions, the EDMs of 
atoms are evaluated by expressing the wave functions as linear combinations of wave functions of states of opposite parities using the first-order perturbation theory. 
In the RNCC method, this applies to both the ket and bra states. Furthermore, one key point in performing the atomic calculations is that the two-body interactions are expanded
in terms of multipoles and their matrix elements require dealing with spherical tensors. Therefore, introducing additional operators in the RNCC method is
computationally very expensive, but it can be handled efficiently by modern day supercomputers. In view of the steady advances in the EDM experiments on diamagnetic atoms in recent
years \cite{graner,parker}, it is essential to develop theories like the RNCC method for improving the accuracy of the atomic calculations in order to probe BSM physics.
 
In this Letter, we outline the general theory of the RNCC method for atoms in the presence of P,T-odd interaction Hamiltonians and the electric dipole operator as external perturbations and 
discuss its implementation for the determination of the EDM ($d_a$) and the electric dipole polarizability ($\alpha_d$) of atomic systems, respectively. As the first application, we 
evaluate these properties for $^{199}$Hg using the RNCC method and compare them with the results from the RCC method, and also the latter property is compared with its measured 
value. 

We begin with the Dirac-Coulomb Hamiltonian, which is given in atomic units by
\begin{eqnarray}
H_a &=&\sum_i  [ c\mbox{\boldmath$\alpha$}_D\cdot \textbf{p}_i+(\beta_D -1)c^2+
V_N(r_i) + \sum_{j \ge i} \frac{1}{r_{ij}} ], \ \ \ \ \
\end{eqnarray}
for $\mbox{\boldmath$\alpha$}_D$ and $\beta_D$ are  the Dirac matrices, $c$ is the velocity of light, and $V_N(r)$ is the nuclear potential experienced
by the electrons in an atom. 

The P,T-odd T-PT interaction Hamiltonian is given by \cite{yamanaka}
\begin{eqnarray}
H_{e-N}^{TPT} = i \sqrt{2} G_F C_T \sum_e \mbox{\boldmath $\sigma_N \cdot \gamma$} \rho_N(r) , 
 \label{htpt}
\end{eqnarray}
where $G_F$ is the Fermi constant, $C_T$ is the T-PT e-N coupling constant, {\boldmath$\sigma_N$}$=\langle \sigma_N \rangle {\bf I}/I$ is the Pauli spinor for the 
nucleus with spin $I$ and $\rho_N(r)$ is the nuclear density. 

The P,T-odd Hamiltonian representing the interaction of the NSM with an electron in an atom is given by \cite{yamanaka}
 \begin{eqnarray}
  H_{e-N}^{NSM}= \frac{3{\bf S.r}}{B_4} \rho_N(r),
 \end{eqnarray}
where ${\bf S}=S \frac{{\bf I}}{I}$ is the NSM and $B_4=\int_0^{\infty} dr r^4 \rho_N(r)$. 

A convenient starting point for the ground state wave function of a closed-shell atom like  $^{199}$Hg is the mean-field Dirac-Hartree-Fock (DHF) wave function ($|\Phi_0 \rangle$). Taking $|\Phi_0 \rangle$ as the reference state, the ground state wave function ($|\Psi_0^{(0)} \rangle$) is expressed in the RCC method 
as \cite{cizek} 
\begin{eqnarray}
 |\Psi_0^{(0)} \rangle = e^{T^{(0)}} |\Phi_0 \rangle,
 \label{eqt0}
\end{eqnarray}
where $T^{(0)}$ is the even parity RCC excitation operator due to the residual Coulomb interaction; i.e. the difference of the two-body Coulomb and the DHF
potential energies. In the particle-hole excitation formalism, we express $T^{(0)}=\sum_{I=1}^{N_c} t_I^{(0)} C_I^{+}$ with $N_c$ representing number of electrons in the system, 
$t_I^{(0)}$ are the amplitudes of the excitations and $C_I^{+}$ stands for a string of annihilation-creation operators corresponding to a general particle-hole excitation. The 
equation for the ground state of $H_a$ is given by
\begin{eqnarray}
 H_a |\Psi_0^{(0)} \rangle &=& E_0^{(0)} |\Psi_0^{(0)} \rangle
 \label{eqm0}
\end{eqnarray}
with energy $E_0^{(0)}$. The equations for the cluster amplitudes $T^{(0)}$ are obtained by using Eq. (\ref{eqt0}) and projecting Eq. (\ref{eqm0}) on the bra state
$\langle \Phi_0| C_I^{-}e^{-T^{(0)}}$ as 
\begin{eqnarray}
\langle \Phi_0| C_I^{-} \overline{H}_a|\Phi_0\rangle=0,  \label{eq36} 
\end{eqnarray}
where the de-excitation operators $C_I^{-}$ are the hermitian conjugate (h.c.) of $C_I^{+}$. We use the notation $\overline{O}=e^{-T}Oe^T=(Oe^T)_{c}$ through out the paper for a 
general operator $O$, where the subscript $c$ stands for connected terms \cite{bartlett}. For one-body and two-body operators $O$, $\overline{O}$ terminates naturally 
\cite{bartlett,crawford}. Also in the case of the one-body operator $O$, it consists of only a few terms.

The inclusion of a weak P,T-odd interaction Hamiltonian or the electric dipole operator $D$, denoted by $H_{\lambda}$, will modify the ground state wave function which can be 
written as 
\begin{eqnarray}
 |\Psi_0 \rangle = e^T |\Phi_0 \rangle=e^{T^{(0)}+\lambda T^{(1)}} |\Phi_0 \rangle,
\end{eqnarray}
where the effect of the perturbation is represented by $T^{(1)}=\sum_{I=1}^{N_c} t_I^{(1)} C_I^{+}$ with the amplitudes $t_I^{(1)}$, which includes one-order of the weak odd-parity
perturbation of interest and all-orders of the residual Coulomb interaction. Here $\lambda$ represents the strength of the coupling coefficient of a given
P,T-odd interaction or the electric field for the evaluation of $d_a$ and $\alpha_d$, respectively. $|\Psi_0\rangle$ is clearly a mixed parity state and to the first-order 
in one of the odd-parity operators, we can express \cite{yamanaka,yashpal,bksahoo}
\begin{eqnarray}
 |\Psi_0 \rangle \simeq |\Psi_0^{(0)} \rangle + \lambda |\Psi_0^{(1)} \rangle.
\end{eqnarray}
This corresponds to
\begin{eqnarray}
 |\Psi_0^{(1)} \rangle = e^{T^{(0)}} T^{(1)} |\Phi_0 \rangle.
 \label{eqt1}
\end{eqnarray}
The first-order perturbed wave function satisfies the following equation
\begin{eqnarray}
 (H_a - E_0^{(0)})|\Psi_0^{(1)} \rangle = (E_0^{(1)} - H_{\lambda}) |\Psi_0^{(0)} \rangle
\end{eqnarray}
with the first-order perturbed energy $E_0^{(1)} = 0$,  since the unperturbed atomic states have a definite parity. The equation for the amplitudes for $T^{(1)}$ can be obtained
from the above first-order perturbed equation as \cite{yamanaka,yashpal,bksahoo}
\begin{eqnarray}
\langle \Phi_0|C_I^{-} \left [ \overline{H}_aT^{(1)} + \overline{H}_{\lambda}\right ] |\Phi_0\rangle =0 . \label{eq37} \  \
\end{eqnarray}
For the calculations of $|\Psi_0^{(0)} \rangle$ and $|\Psi_0^{(1)} \rangle$, we consider singles (one particle-one hole) and doubles (two particle-two hole) excitations 
in the RCC theory (CCSD method) by restricting to $I=1,2$ in the amplitude equations. The expectation value of an operator $O$ in the ground state of a closed-shell system using 
the (R)CC method is expressed as \cite{cizek,bartlett,bishopbook}
\begin{eqnarray}
< O > &\equiv& \frac{\langle \Psi_0 | O | \Psi_0 \rangle}{\langle \Psi_0 | \Psi_0 \rangle} =  \frac{\langle\Phi_0 | e^{T^{\dagger}} O e^T | \Phi_0 \rangle }
                  {\langle\Phi_0 | e^{T^{\dagger}} e^T | \Phi_0 \rangle } \nonumber \\
                  &=& \langle\Phi_0 | e^{T^{\dagger}} O e^T | \Phi_0 \rangle_{c} .     
\label{eqcco}
\end{eqnarray}

\begin{table}[t]
\caption{A summary of $d_a$ values from the T-PT e-N interaction (in $10^{-20}C_T \langle \sigma_N\rangle$ $|e|$cm) and NSM (in ($10^{-17}[S/|e|fm^3]$ $|e|$cm) 
and $\alpha_d$ (in $|e|a_0^2$) in the $^{199}$Hg atom from different methods that are discussed in Refs. \cite{yamanaka,bksahoo,yashpal} and NCCSD method. 
Here CCSD$^{(\infty)}$ values are the final CCSD results.} 
\begin{ruledtabular}
\begin{tabular}{lccc}
Method & T-PT  & NSM  & $\alpha_d$ \\
\hline 
   &  &  & \\
DHF & $-$2.39 & $-$1.20 & 40.95 \\
MBPT(2) & $-4.48$ & $-2.30$ & 34.18 \\
MBPT(3) & $-3.33$ & $-1.72$ & 22.98 \\
RPA     & $-5.89$ & $-2.94$ & 44.98 \\
CI$+$MBPT & $-5.1$ & $-2.6$ & 32.99 \\
MCDF & $-4.84$ & $-2.22$ & \\
PRCC & $-4.3$ & $-2.46$ & 33.29 \\  
LCCSD  & $-4.52$  &  $-2.24$ & 33.91 \\
CCSD$^{(2)}$ & $-3.82$ & $-2.00$ & 33.76 \\
CCSD$^{(4)}$ & $-4.14$ & $-2.05$ & 35.13 \\
CCSD$^{(5)}$ & $-4.02$ & $-2.00$ & 34.98 \\
CCSD$^{(\infty)}$ & $-3.17$ & $-1.76$ & 34.51 \\
NCCSD & $-3.30$ & $-1.77$ & 34.22 \\    
& & & \\
Experiment \cite{goebel} & &  & 33.91(34) \\
\end{tabular}
\end{ruledtabular}
\label{tab1}
\end{table}

Following the above expression and expanding $|\Psi_0 \rangle$, we can evaluate $d_a$ and $\alpha_d$, commonly denoted as $X$, of the ground state of a closed-shell atom by \cite{yashpal}
\begin{eqnarray}
X & \equiv & \lambda \langle \Psi_0^{(1)} | D | \Psi_0^{(0)} \rangle + \langle \Psi_0^{(0)} | D | \Psi_0^{(1)} \rangle \nonumber \\ 
                  &=&  \lambda \langle\Phi_0 | e^{T^{(0) \dagger}} D e^{T^{(0)}} T^{(1)} + T^{(1)\dagger} e^{T^{(0)\dagger}} D e^{T^{(0)}} | \Phi_0 \rangle_{cc} \nonumber \\
                  &=& 2 \lambda \langle\Phi_0 | e^{T^{(0)\dagger}} D e^{T^{(0)}} T^{(1)} | \Phi_0 \rangle_{c} . \ \ \ \label{eqccx}
\end{eqnarray}
The factor 2 appears as the above RCC terms are equal to their h.c. terms. The above expression is non-terminating because of $e^{T^{\dagger (0)}} D e^{T^{(0)}}$. The normalization
factor will also cancel out from the numerator and the denominator in the expectation value when the odd parity Hamiltonian is added to the atomic Hamiltonian. The atomic wave 
function in the expectation value will contain the coupling parameter of the odd parity interaction to all-orders, but as mentioned earlier it is sufficient to consider only one
of this quantity in the actual calculations. Unlike this approach, the normalization factor appears explicitly in the PRCC method \cite{latha,mani,angom}. It is, therefore, necessary 
to adopt a method that can overcome the non-terminating terms and resolve the ambiguity of accounting for contributions from the normalization of the wave function. As
discussed below the RNCC method achieves both these objectives in a natural manner. 

In the RNCC method, the unperturbed ket is the same as that in the RCC method, but the bra $\langle \Psi_0^{(0)} |$ is replaced by $\langle \tilde{\Psi}_0^{(0)} |$ and is defined as 
\cite{bishop,bishopbook1,bishopbook,arponen}
\begin{eqnarray}
 \langle \tilde{\Psi}_0^{(0)} | = \langle \Phi_0 | (1+\tilde{T}^{(0)}) e^{-T^{(0)}} ,
 \label{eqbr}
\end{eqnarray}
where $\tilde{T}^{(0)}=\sum_{I=1}^{N_c} \tilde{t}_I^{(0)} C_I^{-}$ is an de-excitation operator with amplitudes $\tilde{t}_I^{(0)}$, similar to 
$T^{\dagger(0)}=\sum_{I=1}^{N_c} t_I^{(0)} C_I^{-}$, such that it satisfies the bi-orthogonal condition 
\begin{eqnarray}
 \langle \tilde{\Psi}_0^{(0)} | \Psi_0^{(0)} \rangle = \langle \Phi_0 | (1+\tilde{T}^{(0)}) e^{-T^{(0)}} e^{T^{(0)}} |\Phi_0 \rangle =1 .
\end{eqnarray}
It can be easily shown that $\langle \tilde{\Psi}_0^{(0)} |$ has the same eigenvalue as $| \Psi_0^{(0)} \rangle$ (or $\langle \Psi_0^{(0)} |$) with one and only condition that
\begin{eqnarray}
\langle \Phi_0 | \tilde{T}^{(0)} \overline{H}_a|\Phi_0\rangle= 0 .
\end{eqnarray}
In fact, this condition is the direct consequence of Eq. (\ref{eq36}). Hence,  $\langle \tilde{\Psi}_0^{(0)} |$ can be used in place of 
$\langle \Psi_0^{(0)} |$ in the calculation of atomic properties. Furthermore, this choice of the bra ensures that the Hellman-Feynman
equation is satisfied \cite{bishopbook1}.

\begin{table}[t]
\caption{Comparison of contributions from various RCC and RNCC terms to $d_a$ and $\alpha_d$ values (with same units as in Table \ref{tab1}). It can be noticed that contributions 
using different bra states in the two methods show very different trends, but the final results are in very good agreement.}
\begin{ruledtabular}
\begin{tabular}{lclc}
 RCC term & RCC result & RNCC term & RNCC result  \\
\hline
 & & & \\
 \multicolumn{4}{c}{Contributions to $d_a$ from T-PT interaction} \\
$DT_1^{(1)}$                & $-2.20$  &  $DT_1^{(1)}$  & $-2.20$ \\
$T_1^{(1)\dagger} D$        & $-2.20$  & $\tilde{T}_1^{(1)} D$ & $-1.74$ \\
$T_1^{(1)\dagger}DT_2^{(0)}$&  0.61 & $\tilde{T}_1^{(1)} DT_2^{(0)}$  & 0.52 \\
$T_2^{(0)\dagger}DT_2^{(1)}$& 0.01  & $\tilde{T}_2^{(0)} DT_2^{(1)}$  & 0.01 \\
$T_2^{(1)\dagger}DT_2^{(0)}$& 0.01  & $\tilde{T}_2^{(1)} DT_2^{(0)}$  & $-0.04$\\
Others & 0.60 & Others & 0.15 \\
\hline 
 & & & \\
 \multicolumn{4}{c}{Contributions to $d_a$ from NSM interaction} \\
$DT_1^{(1)}$                & $-1.19$  &  $DT_1^{(1)}$  & $-1.19$ \\
$T_1^{(1)\dagger} D$        & $-1.19$  & $\tilde{T}_1^{(1)} D$ & $-0.88$ \\
$T_1^{(1)\dagger}DT_2^{(0)}$& 0.30 & $\tilde{T}_1^{(1)} DT_2^{(0)}$  & 0.25 \\
$T_2^{(0)\dagger}DT_2^{(1)}$& $-0.01$  & $\tilde{T}_2^{(0)} DT_2^{(1)}$  & $-0.01$ \\
$T_2^{(1)\dagger}DT_2^{(0)}$& $-0.01$  & $\tilde{T}_2^{(1)} DT_2^{(0)}$  & $-0.02$\\
Others & 0.34 & Others & 0.08 \\
\hline 
 & & & \\
 \multicolumn{4}{c}{Contributions to $\alpha_d$} \\
$DT_1^{(1)}$                & 20.43  &  $DT_1^{(1)}$  & 20.43 \\
$T_1^{(1)\dagger} D$        & 20.43  & $\tilde{T}_1^{(1)} D$ & 16.72 \\
$T_1^{(1)\dagger}DT_2^{(0)}$&  $-2.93$ & $\tilde{T}_1^{(1)} DT_2^{(0)}$  & $-2.42$ \\
$T_2^{(0)\dagger}DT_2^{(1)}$& 0.71  & $\tilde{T}_2^{(0)} DT_2^{(1)}$  & 0.67\\
$T_2^{(1)\dagger}DT_2^{(0)}$& 0.71  & $\tilde{T}_2^{(1)} DT_2^{(0)}$  & 0.68\\
Others & $-4.84$ & Others & $-1.86$ \\
\end{tabular} 
\end{ruledtabular}
\label{tab2}
\end{table}

Starting from the bra equation  
\begin{eqnarray}
 \langle \tilde{\Psi}_0^{(0)} | H_a &=& E_0^{(0)} \langle \tilde{\Psi}_0^{(0)} |,
 \label{eqmn}
\end{eqnarray}
the amplitudes of $\tilde{T}^{(0)}$ can be determined using Eq. (\ref{eqbr}) and projecting Eq. (\ref{eqmn}) on the ket state  $e^{T^{(0)}} C_I^{+} |\Phi_0 \rangle$ as
\begin{eqnarray}
\langle \Phi_0|\left [ \tilde{T}^{(0)} \overline{H}_a + \overline{H}_a\right ] C_I^{+} |\Phi_0 \rangle =0 .
\label{eq39}
\end{eqnarray}
It is interesting to note that the ket and bra equations, Eqs. (\ref{eq36}) and (\ref{eq39}) respectively, can be derived from a variational principle \cite{arponen}. After obtaining these
solutions, the expectation value of an operator $O$ in the RNCC method can be written as \cite{bishop,bishopbook1,bishopbook}
\begin{eqnarray}
< O > &\equiv& \frac{\langle \tilde{\Psi}_0^{(0)} | O | \Psi_0^{(0)} \rangle}{\langle \tilde{\Psi}_0^{(0)} | \Psi_0^{(0)} \rangle} = 
\langle\Phi_0 | (1+ \tilde{T}^{(0)}) \overline{O} | \Phi_0 \rangle .
\end{eqnarray}
The above expression, unlike its counterpart in the RCC method, is a terminating series and the normalization factor is unity owing to the aforementioned bi-orthogonal condition. 
But to evaluate EDMs and electric dipole polarizabilities of atoms, the above (R)NCC approach needs further modification as demonstrated in this work.

To obtain the first-order correction to the bra state, we can replace $\langle \Psi_0|$ of Eq. (\ref{eqcco}) by $\langle \tilde{\Psi}_0 |$ defining as
\begin{eqnarray}
 \langle \tilde{\Psi}_0 | &=& \langle \Phi_0 | (1+\tilde{T}) e^{-T}
\end{eqnarray}
where $\tilde{T}$ is the new de-excitation RCC operator for the total wave function. Expanding the bra wave function and retaining terms to the first-order yields
\begin{eqnarray}
\langle \tilde{\Psi}_0 | &\simeq& \langle \tilde{\Psi}_0^{(0)} | + \lambda \langle \tilde{\Psi}_0^{(1)} | \nonumber \\
 &=& \langle \Phi_0 | (1+\tilde{T}^{(0)} + \lambda \tilde{T}^{(1)}) e^{-(T^{{0}} + \lambda T^{(1)})} . 
\end{eqnarray}
It is straightforward to show that $\langle \tilde{\Psi}_0 |$ can satisfy the bi-orthogonal condition with $|\Psi_0 \rangle$. Analogous to Eq. (\ref{eq37}), the amplitudes for
$\tilde{T}^{(1)}$ starting from the first-order perturbed bra can be obtained by solving
{\small{
\begin{eqnarray}
\langle \Phi_0 | \left [ \tilde{T}^{(1)} \overline{H}_a  + (1+ \tilde{T}^{(0)}) \left \{ \overline{H}_{\lambda}+ (\overline{H}_aT^{(1)})_{c} \right \} \right ] C_I^{+}|\Phi_0 \rangle =0. \ \ \ \ \
\end{eqnarray}
}}
Again, we express  $\tilde{T}^{(0/1)}=\tilde{T}_1^{(0/1)}+\tilde{T}_2^{(0/1)}$ in the singles and doubles approximation of the RNCC method (NCCSD method) corresponding to their 
respective $T^{(0/1)}$ operators in the CCSD method. In this approach, the expectation value $X$ is evaluated by
\begin{eqnarray}
X &\equiv& \frac{\langle \tilde{\Psi}_0 | D | \Psi_0 \rangle}{\langle \tilde{\Psi}_0 | \Psi_0 \rangle} =  \langle\Phi_0 | (1+\tilde{T}) e^{-T} D e^T | \Phi_0 \rangle  \nonumber \\
                  &=& \lambda \langle\Phi_0 | (1+\tilde{T}^{(0)})  \overline{D} T^{(1)} + \tilde{T}^{(1)} \overline{D} | \Phi_0 \rangle . 
\label{eqnccx}
\end{eqnarray}
This expression terminates and can give fewer terms than Eq. (\ref{eqccx}). Also, it does not have any h.c. terms like those in the RCC method. Thus, one to one comparison 
between the contributions from various terms from Eqs. (\ref{eqccx}) and (\ref{eqnccx}) will be instructive. Substantial discrepancies between the final results of 
the RNCC and the RCC methods will reflect the incompleteness of the latter.

We give a summary of the results obtained for $d_a$ considering both the T-PT and NSM P,T-odd interaction Hamiltonians and also $\alpha_d$ value of $^{199}$Hg using the DHF, 
many-body perturbation theory with second (MBPT(2)) and third-order (MBPT(3)) approximations, random-phase approximation (RPA), combined configuration interaction
and many-body theory (CI$+$MBPT), multi-configuration Dirac-Fock (MCDF), PRCC and RCC approaches, that were extensively discussed recently in Refs. \cite{yamanaka,bksahoo,yashpal},
in Table \ref{tab1}. We also quote values explicitly from the linear terms of the CCSD (LCCSD) method and a self-consistent CCSD approach in which the combined power of $T^{(0)}$
and $T^{(0)\dagger}$ is systematically increased in Eq. (\ref{eqccx}) till the value of the EDM converges. The latter method is designated as CCSD$^{(k)}$ for $k$ number of $T^{(0)}$ 
and $T^{(0)\dagger}$ operators in the non-terminating series. It is evident from Table \ref{tab1} that there are discrepancies between the results based on different methods. The results 
from CCSD$^{(\infty)}$ and NCCSD are in very good agreement as expected, but they differ significantly from those of other methods. This demonstrates the importance of the correlation
effects embodied in the higher-order particle-hole excitations that are present in the two aforementioned CC methods but not in the CI$+$MBPT and the MCDF calculations. Another 
possible reason for the disagreement could be that unlike the CC methods the latter two methods are not size-extensive. The $\alpha_d$ value from these methods are also compared
with the experimental result \cite{goebel}. As can be seen from the table, the polarizability result is slightly closer to the central experimental value and the EDM results for both
the T-PT and NSM interactions are slightly larger for the NCCSD method compared to those obtained using the CCSD method. In Table \ref{tab2}, we also compare contributions to $d_a$ 
for both the T-PT and NSM interactions and  $\alpha_d$  from individual terms in the CCSD and NCCSD methods. It clearly shows that there are large differences in the contributions 
between the h.c. terms of the RCC method and their counterparts in the RNCC method. However, the differences in the final results for the two methods given in Table \ref{tab1} are
negligibly small. Thus the close agreement between the two results validate our results for both the CCSD (with $k=\infty$) and NCCSD methods. In view of this and the high accuracy
of our polarizability calculations mentioned above, we estimate that the error in these two CC calculations due to basis sets and neglected correlation effects will 
not exceed 2\%.

 Combining our calculations from the NCCSD method with the upper limit of EDM in the $^{199}$Hg atom $|d_a( ^{199}{\rm{Hg}})|< 7.4\times 10^{-30} |e|\rm{cm}$ \cite{graner}, we get  
\begin{eqnarray}
 |C_T| < \ 6.73 \times 10^{-10} 
\end{eqnarray}
and 
\begin{eqnarray}
 |S| < 4.18 \times 10^{-13} |e| \rm{fm}^3 ,
 \label{eqls}
\end{eqnarray}
which are very close to the values reported by us earlier \cite{bksahoo,yamanaka}. Therefore, this suggests that limits obtained for 
$|\bar{\theta}|$ and $|\tilde{d}_u - \tilde{d}_d|$ in Refs. \cite{bksahoo,yamanaka} are very reliable.

In  conclusion, we have presented a general theory of the RNCC method incorporating the P,T-odd interactions  and the electric dipole operator as perturbations to determine the
EDMs and the electric dipole polarizability of closed-shell atomic systems, respectively. This theory has been applied to evaluate the aforementioned properties of $^{199}$Hg 
for which the EDM limit is the lowest to date. The polarizability using the RNCC method has been obtained within the error bar of the precisely measured value of this quantity. 
The RNCC results are also in very good agreement with the EDMs and the polarizability that have been determined using a self-consistent procedure based on the RCC method, which
differs from earlier calculations by 20-50\%. These findings validate the previous self-consistent RCC calculations of EDMs in $^{199}$Hg that were 
used to extract accurate limits on various fundamental CP violating parameters for nuclear and particle physics. It is also imperative to carry out similar 
analyses by applying the RNCC method to evaluate the EDMs of $^{171}$Yb and $^{225}$Ra, where the differences between the calculations have been found to be even 
larger due to the strong electron correlation effects in these atoms. This method will clearly be a valuable tool for the accurate theoretical determination of the EDMs and 
electric dipole polarizabilities of closed-shell atoms.

B. K. S. acknowledges financial support from Chinese Academy of Science (CAS) through the PIFI fellowship under the project number 2017VMB0023. Computations
were carried out using Vikram-100 HPC cluster of Physical Research Laboratory (PRL), Ahmedabad, India.


\begin{thebibliography}{50}

\bibitem{barr}
S. M. Barr, Int. J. Mod. Phys. A {\bf 8}, 209 (1993).

\bibitem{pospelovreview}
M. Pospelov and A. Ritz, Ann. Phys. (N.Y.) {\bf 318}, 119 (2005).

\bibitem{graner}
B. Graner, Y. Chen, E. G. Lindahl, and B. R. Heckel, Phys. Rev. Lett. {\bf 116}, 161601 (2016).

\bibitem{yamanaka}
N. Yamanaka {\it et al}, Eur. Phys. J. A {\bf 53}, 54 (2017).

\bibitem{yashpal}
Y. Singh and B. K. Sahoo, Phys. Rev. A {\bf 91}, 030501(R) (2015).

\bibitem{bksahoo}
B. K. Sahoo, Phys. Rev. D {\bf 95}, 013002 (2017).

\bibitem{latha}
K. V. P. Latha {\it et al}, Phys. Rev. Lett. {\bf 103}, 083001 (2009); Erratum: Phys. Rev. Lett. {\bf 115}, 059902 (2015).

\bibitem{mani}
S. Chattopadhyay, B. K. Mani and D. Angom, Phys. Rev. A {\bf 91}, 052504 (2015).

\bibitem{angom}
D. Angom, {\it Private communication}.

\bibitem{cizek}
J. Cizek, Adv. Chem. Phys. {\bf 14}, 35 (1969).

\bibitem{bartlett}
I. Shavitt and R. J. Bartlett, {\it Many-body methods in Chemistry and Physics}, Cambidge University Press, Cambridge, UK (2009).

\bibitem{sahoo}
B. K. Sahoo and Y. Singh, Phys. Rev. A {\bf 95}, 062514 (2017).

\bibitem{singh}
Y. Singh and B. K. Sahoo, Phys. Rev. A {\bf 92}, 022502 (2015).

\bibitem{deyonker}
N. J. DeYonker and K. A. Peterson, J. Chem. Phys. {\bf 138}, 164312 (2013).

\bibitem{nataraj}
H. S. Nataraj {\it et al.}, Phys. Rev. Lett. {\bf 101}, 033002 (2008).

\bibitem{pathak}
H. Pathak {\it et al.}, Phys. Rev. A {\bf 90}, 010501(R) (2014).

\bibitem{eliav}
L. F. Pasteka, E. Eliav, A. Borschevsky, U. Kaldor and P. Schwerdtfeger, Phys. Rev. Lett. {\bf 118}, 023002 (2017).

\bibitem{crawford}
T. D. Crawford and H. F. Schaefer, Rev. Comp. Chem. {\bf 14}, 33 (2000).

\bibitem{prasana}
V. S. Prasannaa {\it et al.}, Phys. Rev. Lett. {\bf 114}, 183001 (2015).

\bibitem{bishop01}
R. F. Bishop and P. H. Y. Li, Phys. Rev. B {\bf 95}, 134414 (2017).

\bibitem{bishop02}
R. F. Bishop, R. G. Hale, and Y. Xian, Phys. Rev. Lett. {\bf 73}, 3157 (1994).

\bibitem{bishop}
R. F. Bishop, Theoretica Chimica Acta {\bf 80}, 95 (1991).

\bibitem{kowalski}
K. Kowalski, D. J. Dean, M. Hjorth-Jensen, T. Papenbrock, and P. Piecuch, Phys. Rev. Lett. {\bf 92}, 132501 (2004).

\bibitem{hagen}
G. Hagen, T. Papenbrock, M. Hjorth-Jensen, and D. J. Dean, Rep. Prog. Phys. {\bf 77}, 096302 (2014).

\bibitem{lindgren}
I. Lindgren and J. Morrison, {\it Atomic Many-Body Theory}, Second Edition, Springer-Verlag, Berlin, Germany (1986).

\bibitem{bishopbook1}
R. Bishop, J. Arponen and P. Pajanee, {\it Aspects of Many-body Effects in Molecules and Extended Systems}, Berlin: Springer-Verlag (1989).

\bibitem{bishopbook}
R. F. Bishop, {\it Microscopic Quantum Many-Body Theories and their Applications}, Lecture Series in Physics, pg. 1, Springer Publication, Berlin (1998).

\bibitem{bishop1}
R. F. Bishop and J. S. Arponen, Int. J. Quantum Chem. Symp. {\bf 24}, 197 (1990). 

\bibitem{arponen}
J. S. Arponen, Ann. Phys. {\bf 151}, 311 (1983).

\bibitem{parker}
R. H. Parker {\it et al}, Phys. Rev. Lett. {\bf 114}, 233002 (2015). 

\bibitem{goebel}
D. Goebel and U. Hohm, J. Phys. Chem. {\bf 100}, 7710 (1996).

\end{thebibliography}
\end{document}